# Modeling Terrorist Networks - Complex Systems at the Mid-Range[i]

by

**Philip Vos Fellman, American Military University**
Shirogitsune99@yahoo.com

**Roxana Wright, Plymouth State University**
Rox_wright@yahoo.com

---

[i] © Philip Vos Fellman and Roxana Wright, May 14, 2003, Jamuary, 2004, July, 2008.

*"The wrath of the terrorist is rarely uncontrolled. Contrary to both popular belief and media depiction, most terrorism is neither crazed nor capricious. Rather, terrorist attacks are generally as carefully planned as they are premeditated."*

                         Bruce Hoffman, RAND Corporation

*"The best method to control something is to understand how it works."*

                         J. Doyne Farmer, Santa Fe Institute

## **Introduction**

As we started thinking about how to organize this paper, I was reminded of two statements which powerfully impressed me as a student of National Security Policy at Yale University nearly two decades ago. Paul Bracken (1984), who had just recently left the Hudson Institute, where had worked a number of years for Hermann Kahn, the "father of modern nuclear strategy" impressed upon us very early on that National security policy, which in those days was primarily focused upon the command and control of nuclear forces, is characterized by ***irreducible levels of ambiguity and complexity***.

The mathematical revolution ***of chaos theory*** and ***complexity science*** has given us powerful new modeling tools that were unthinkable just a generation ago. ***The pace of technological change has matched the emergence of new sciences*** in ways which likewise would have been difficult to conceive of in even the relatively recent past. For example, one of the primary purposes of the U.S. 1979 Export Administration Act was to prevent the migration of dual-use technologies like the 32 bit architecture of the Intel 80486 Microprocessor, which could be used as a targeting system for ICBM's. Today there's more science and technology involved in controlling the scaling dynamics of internet traffic packet delay than there is in designing the navigational system for mid-course correction on any ballistic missile re-entry system.[ii] When these kinds of technological developments are combined with the new power of ***autonomous non-state actors and various persistent vulnerabilities of complex, self-organizing systems*** ("Avoiding Complexity Catastrophe", McKelvey, 1999) ***the challenges of national security in the 21st century truly take on an entirely different character and require, tools, techniques, resources, models and knowledge which are fundamentally different from their 20th century predecessors***.

In the context of this newly emerging dynamics, a proper approach to modeling terrorist networks and their flows of information, money, and material needs to be

---

[ii] See Qong Li and David Mills, "Investigating the Scaling Behavior, Crossover and Antipersistence

structured in such a way that at the *mid-range*, various government agencies can efficiently share information, spread and reduce risk, especially risk to sensitive infrastructure or epidemiological risk from bio-weapons.

**Terrorism is not Random**

J. Doyne Farmer, of the Santa Fe Institute, captures Hoffman's non-randomness argument an *Edge* (http://www.edge.com) interview when he notes that:[iii]

> Randomness and determinism are the poles that define the extremes in any assignment of causality. Of course reality is usually somewhere in between. Following Poincaré, we say that something is random if the cause seems to have little to do with the effect. Even though there is nothing more deterministic than celestial mechanics, if someone gets hit in the head by a meteor, we say this is bad luck, a random event, because their head and the meteor had little to do with each other. Nobody threw the meteor, and it could just as well have hit someone else. The corresponding point of view here is that bin Laden and his associates are an anomaly, and the fact that they are picking on us is just bad luck. We haven't done anything wrong and there is no reason to change our behavior; if we can just get rid of them, the problem will disappear. This is the view that we would all rather believe because the remedy is much easier.

Farmer goes on to explain the obvious, where he is in substantial agreement with Fuller and Hoffman that while we might like to believe in the "bad luck" theory, terrorism generally has deep, underlying causes and is not likely to go away on its own. In fact, much as Bruce Russet and Paul Kennedy at Yale University have shown in the case of war casualties, which have risen by an order of magnitude each century,[iv] there appears to be an emerging pattern where *the overall number of casualties resulting from terrorism is also growing at exponential rate*.

Yet, getting at the root causes of terrorism is one of those things that falls into the category of *irreducible complexity and ambiguity*. It is, in fact, the very difficulty of the enterprise which leads us towards looking at solutions at the mid-range rather than proposing some hypothetical system or methodology which would render terrorist acts either highly predictable (and hence, theoretically avoidable) or which would allow the dismantling of terrorist organizations as soon as they form. In terms of formal properties of the system, *terrorist behavior* falls somewhere between the purely chaotic and the fully deterministic realms, which we represent as a *non-linear dynamical system, characterized by a low-order chaotic attractor*.

---

[iii] J. Doyne Farmer, "What Now?", http://www.edge.org/documents/whatnow/whatnow_farmer.html
[iv] (a) Bruce Russett, Harvey Starr, David Kinsella, World Politics: The Menu for Choice, Wadsworth Publishing, 6th edition, 1999; (b) Paul M. Kennedy, The Rise and Fall of the Great Powers: Economic Change and Military Conflict from 1500 to 2000, Vintage Books, 1989.

As a pattern of behaviors, terrorism can be modeled in the same way as other phenomena which exhibit regularity but not periodicity (i.e., locally random, but globally defined).[v] Farmer, for example describes the two principal approaches to dealing with prediction in a "chaotic" system. The first is a ***formal predictive methodology***. Relating terrorism to "simple" systems[vi] such as roulette wheels, turbulent fluids and stock markets, he explains:[vii]

> To predict the trajectory of something, you have to understand all the details and keep track of every little thing. This is like solving terrorism by surveillance and security. Put a system in place that will detect and track every terrorist and prevent them from acting. ***This is a tempting solution, because it is easy to build a political consensus for it***, and *it involves technology, which is something we are good at*. But if there is one thing I have learned in my twenty five years of trying to predict chaotic systems, it is this: ***It is really hard, and it is fundamentally impossible to do it well***. This is particularly so when it involves a large number of independent actors, each of which is difficult to predict. We should think carefully about similar situations, such as the drug war: As long as people are willing to pay a lot of money for drugs, no matter how hard we try to stop them, drugs will be produced, and smugglers and dealers will figure out how to avoid interception. We have been fighting the drug war for more than thirty years, and have made essentially no progress. If we take the same approach against terrorism we are sure to fail, for the same reasons.

For this reason, along with the reasons cited in our references, we do acknowledge the impossibility of total predictivity. However, we also still believe that over and above any which might be taken at the state level, the greatest room for improving the performance of those organizations tasked with preventing or combating terrorism is at the ***mid-range***. That is, we think the application of the most recent advances in science is most likely to

---

[v] At a basic mathematical level, this kind of phenomenon is explained very clearly by Edgar Peters in "Chaos and Order in the Capital Markets", John Wiley and Sons, 1992. A more rigorous treatment can be found in "Statistical Mechanics of Complex Networks" by Reka Albert and Laszlo Barbási, arXiv:cond-mat/0106096v1 6Jun2001, http://www.nd.edu/~networks/Papers/review.pdf

[vi] Here, Farmer is having a bit of a laugh at the expense of his audience. By "simple" systems he means complex, non-linear systems whose strange attractor is one of sufficiently low dimension that there is an observable phenomenon of closely packed state space and mapping of the phase space with Lyapunov exponents is relatively tractable. A "complex" system in this context would be one with a sparsely populated state space, with bifurcations taking place so frequently that even if there is a strange attractor, the "curse of dimensionality" makes it computationally intractable. The definitive work on the subject is (a) Farmer's paper, "Chaotic Attractors of an Infinite-Dimensional Dynamical System", Physica D, 4 (1982) 366-393. A good representative demonstration of the techniques involved is (b) Shampine and Thompson's "Solving Delay Differential Equations with dde23" available on the world wide web at http://www.cs.runet.edu/~thompson/webddes/tutorial.html#CITEjdf . (c) Stuart Kauffman also draws on Farmer's treatment in "The Structure of Rugged Fitness Landscapes", Chapter Two of The Origins of Order, Oxford University Press, 1993, pp. 33-67.

[vii] Ibid., No. 2

bear fruit in the fight against terrorism not at the level of state leadership, and not at the level of mapping and predicting the behavior of each individual terrorist, but rather at an intermediate or organizational level, which, following the late Theda Skocpol, we characterize as "action at the mid-range".

**September 11 and Network Analysis Models: Simple Distributional Properties**

The first difficulty which the analyst must face in constructing a network analysis of terrorist organizations is the difficulty of building an accurate map. Valdis Krebs, who has used network analysis to provide an extensive analysis of the 9/11 Hijackers network, explains three problems he encountered very early on. Drawing on the work of Malcolm Sparrow, he notes that three problems are likely to plague the social network analyst regardless of context. These are:[viii]

1. Incompleteness - the inevitability of missing nodes and links that the investigators will not uncover.
2. Fuzzy boundaries - the difficulty in deciding who to include and who not to include.
3. Dynamic - these networks are not static, they are always changing.

In addition, there is rather a bit of a paradox in that even using a more sophisticated methodology, such as measuring the strength of ties in terrorist networks, (i.e. vector vs. scalar values) may still not yield a more useful map. One reason for this is that many of the factors which determine the strength of terrorist ties are prior connections, which are not easily measured and which, over the short run, may leave the analyst with an impractically sparse network.[ix]

The sparseness of terrorist networks, however, is also a bit of a two-edged sword. In Krebs' initial mapping (Figure I), he found that the 19 member network had an average path length of 4.75 steps, with some of the hijackers separated by more than 6 steps, and some of the associates over the observable event horizon. Krebs describes this as trading efficiency for secrecy. Another way of describing this process is extreme compartmentation or even "over-compartmentation".

---

[viii] Valdis Krebs, "Uncloaking Terrorist Networks", *First Monday*, http://www.firstmonday.dk/issues/issue7_4/krebs/
[ix] Ibid.

**Figure I: Valdis Krebs' Initial Mapping of the 9/11 Hijackers' Network**

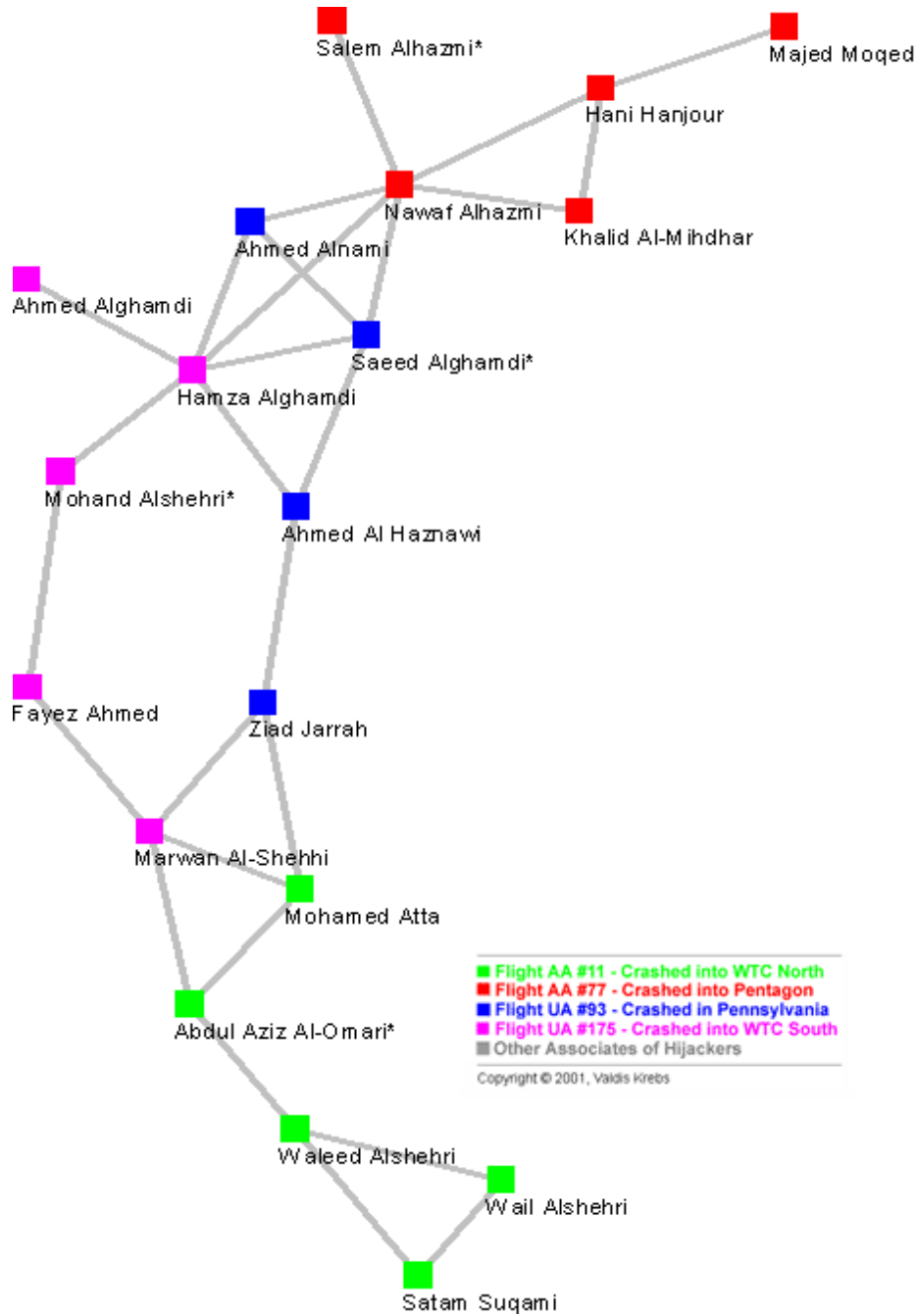

**Complex Networks**: **Beyond Simple Heterogeneous Preferences**

Complex systems tools, especially network analysis, offer some very strong insights about the behavior of terrorist networks, but in the flood of data which has hit the market since 9/11 it is often difficult or impossible to tell who has chosen an appropriate methodology and who has not. Equally difficult is discerning whether a complex model truly possesses the ability to model terrorism in newer, more accurate, more powerful ways. Some models may be mathematically complex, but yield little in the way of practical results, simply because their method is static and terrorist cells are a dynamic phenomenon.[x]

The previously unstructured concept of terrorist networks takes on a new character in the treatment advanced by Krebs.[xi] Krebs' network mapping draws on the application of social network analysis and develops a software system based methodology which he uses to map knowledge networks within and across the boundaries of an organization in order to uncover the ***dynamics of learning and adaptation***. This kind of organizational network analysis combines social network analysis and organizational behavior with chaos theory and complex adaptive systems. The network mapping goes far beyond the formal organizational structure by exposing the real knowledge-sharing dynamics within the functional structures. Krebs describes these ***communities of practice*** as ***emergent groups*** in which knowledge is concentrated around common problems and interests, and the core competencies of an organization are shared and developed (his organizational mapping of terrorist networks derives in large part from his earlier work mapping corporations and studying the dynamics of organizational learning.)

---

[x] A classic mistake in this area is frequently made when the authors apply neoclassical microeconomic "rational actor" assumptions to modeling terrorism which creates a static, homogenous treatment of opponents and sows more confusion than it resolves. While the rational actor methodology was once extremely popular in economics, it has been generally dismissed by "hard" science, and is slowly being replaced by complexity science's heterogeneous agent based modeling. Typically, an agent based model presumes heterogeneous agent composition, preferences and behaviors and uses the *stochastic microagent assumption* to replace the rational actor model. For an explanation of agent-based modeling, see J. Doyne Farmer, "Toward Agent Based Models for Investment", http://www.santafe.edu/~jdf/aimr.pdf , and "Physicists Attempt to Scale the Ivory Towers of Finance", Computing in Science and Engineering, December, 1999, http://www.santafe.edu/sfi/publications/Working-Papers/99-10-073.pdf
For applications of agent based models to terrorism, see Michael Johns and Barry Silverman, "How Emotions and Personality Affect the Utility of Alternative Decisions: A Terrorist Target Selection Case Study" http://www.seas.upenn.edu/~barryg/emotion.pdf , or Ronald A. Woodman, "Agent Based Simulation of Military Operations Other Than War: Small Unit Combat, Thesis, Naval Postgraduate School, Monterey, CA, September, 2000
http://diana.gl.nps.navy.mil/~ahbuss/StudentTheses/WoodamanThesis.pdf
[xi] Valdis Krebs, (a) "Mapping Networks of Terrorist Cells", Connections 24(3): 43-52 http://www.orgnet.com/MappingTerroristNetworks.pdf (b) "Surveillance of Terrorist Networks", http://www.orgnet.com/tnet.html (c) Social Network Analysis of the 9-11 Terrorist Network, http://www.orgnet.com/hijackers.html

The measurement of these "complex human structures" focuses on individual **network centrality**, which reveals key individuals in the information flow and knowledge exchange. **High centrality scores demonstrate extensive access to "hidden assets"** within the organization of an entity with high capacity to "get things done". Network centrality relates the performance of the network as shown below.

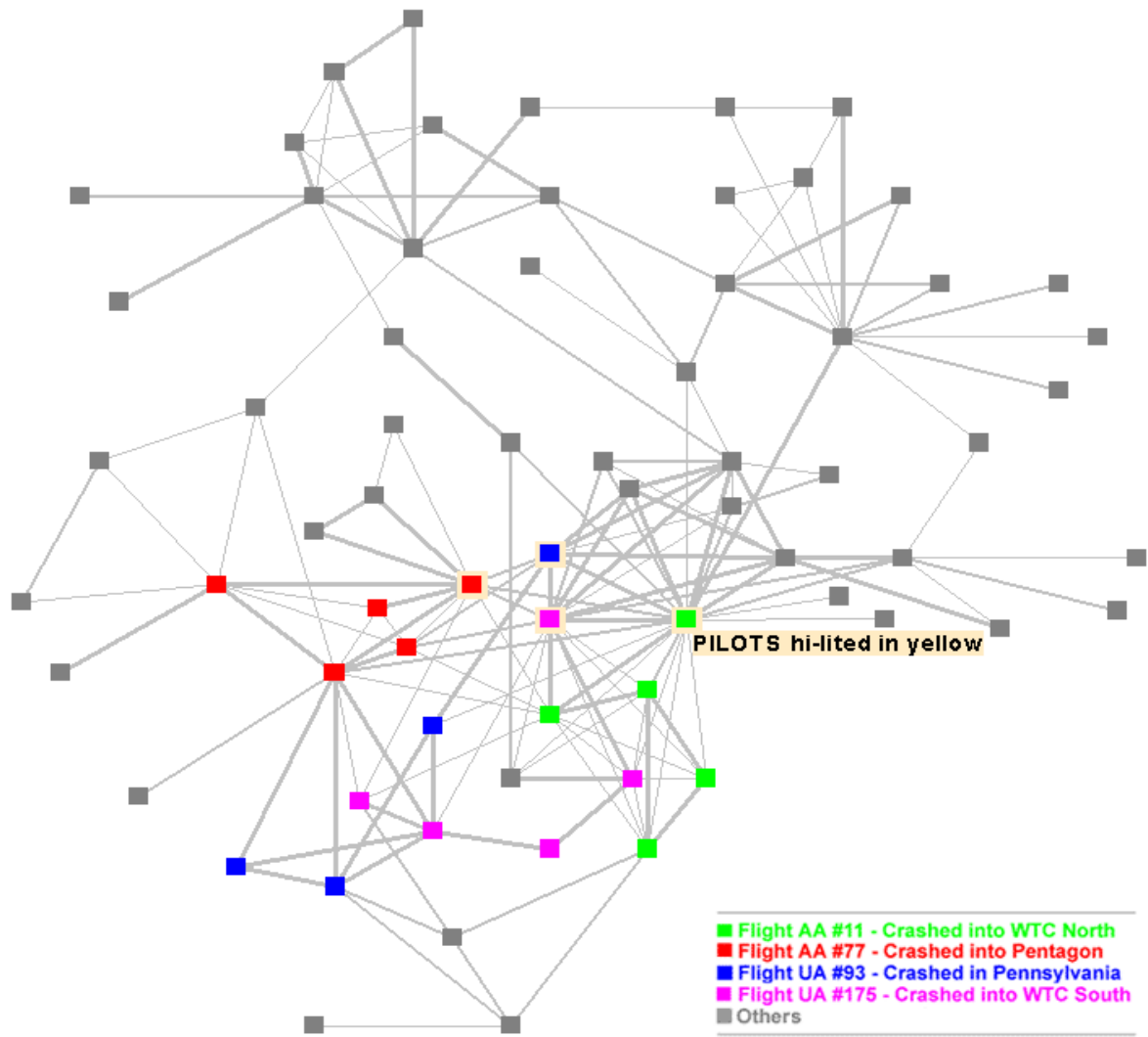

**Figure 2: Krebs' extended model 9-11 hijacker model with measures of centrality**

What is interesting about this second plot is that it illuminates the ways in which terrorist compartmentation dictates the operational parameters of a terrorist attack. Such mappings may also yield previously hidden information about the command structure of terrorist organizations. In "Six Degrees of Mohammed Atta", Thomas Stewart describes several important features of Krebs' network, and points out that:[xii]

> It is not a complete picture; among other problems, it shows only those links that have been publicly disclosed. Still, it's possible to make some interesting inferences. First, the greatest number of lines led to Atta, who scores highest on all three measures, with Al-Shehhi, who is second in both activity and closeness, close behind. However, Nawaf Alhazmi, one of the American Flight 77 hijackers, is an interesting figure. In Krebs's number crunching, Alhazmi comes in second in betweenness, suggesting that he exercised a lot of control, but fourth in activity and only seventh in closeness. But if you eliminate the thinnest links (which also tend to be the most recent -- phone calls and other connections made just before Sept. 11), ***Alhazmi becomes the most powerful node in the net. He is first in both control and access, and second only to Atta in activity. It would be worth exploring the hypothesis that Alhazmi played a large role in planning the attacks, and Atta came to the fore when it was time to carry them out.***

To return to measures of centrality, and the dynamic operational advantage which high centrality incurs, one must first understand how such a system can take advantage of wide degrees of separation between cells. Operational benefits arise from the pattern of connections surrounding a node that allows for wide network reach with minimal direct ties. "Structural holes" at the intersection of flows across knowledge communities position unique and superior nodes. It is the individuals spanning these "*internal holes of opportunity*" that impact the network's functioning and performance. The implicit corollary of this is that if a small number of these critical nodes can be identified and "clipped" from the network, then command signals will not be able to propagate through the system.

In Krebs' mapping, *the main centrality measures are degrees* (number of direct connections that a node has), *betweenness* (the ability of an individual to link to important constituencies) and *closeness* (a position's ability to monitor the information flow and to "see" what is happening in the network). The knowledge flow is facilitated and influenced by boundary spanners with access to information flowing in other clusters, as well as peripheral players that bring fresh information into the network. A network with a low centralization score is more resilient in that it has no single highly-central points of failure. These networks "fail gracefully" as the damage of a node does not lead to a breakdown in information flows and coordination links.

---

[xii] Thomas A. Stewart, "Six Degrees of Mohammed Atta", Business 2.0, December, 2001, http://www.business2.com/articles/mag/0,1640,35253,FF.html

**Social Network Theory**

What social network analysis contributes to counter-terrorism is *the ability to map the invisible dynamics inside a terrorist community*. The methodology draws upon graphical representation in exploring and presenting the patterns displayed by structural data. In the case of terrorist networks, surveillance of the daily activities and contacts of suspects reveals the network around them and thus adds more nodes and links of intentional contacts to the map. Once the direct links are identified, and the "connections of the connections" are included, the key individuals begin to stand out. In 2000, the Central Intelligence Agency identified al-Qaeda suspects Nawaf Alhazmi and Khalid Almihdhar attending a meeting in Malaysia. The mapping of the links between the terrorists involved in the WTC attacks shows that all 19 hijackers were within two degrees from these original suspects, while they also had multiple ties back into the network.

Based on publicly released information from the investigation of the 9/11 terrorists Krebs mapped and evaluated the links which tied the network together and analyzed its resilience. Each link's strength was evaluated based upon the amount of time members spent together. Interactions were rank ordered so that those living together or attending the same training were assigned the strongest ties, the terrorists traveling or participating in common meetings were given ties of moderate strength, and finally, those ties which reflected only occasional relations were characterized as weak links. The thickness of the lines in Figure 2 corresponds to the strength of the ties between the terrorists. This mapping exhibits a dispersed but well-defined structure, although as mentioned earlier, the connections between members are more than usually distant.

As the positioning of Alhazmi discussed above suggests, strong ties may have been inactive and hidden for relatively long intervals, whereas a minimum of weak ties ensured secrecy. This configuration reveals a network which was consciously constructed on the principle of minimizing damage to the organization as a whole in the event that a link is compromised.

This type of network can only achieve its goals by the use of transitory shortcuts that temporarily balance the need for covertness with the need for intense information flow and coordination in active times.[xiii] Sources of public information show that the dense and resilient ties forged in the past were "invisible" during the hijackers' stay in the US. This "massive redundancy through trusted prior contacts" is considered one the major hidden strengths of this network. Such a finding once again highlights the need for human collection, particularly in remote locations. All the tasking in the world by Homeland Security or any other security service won't matter a whit if the ability to identify strong connective linkages between terrorists.

After the 9/11 disaster, there was a trend in Washington to talk about the hijackings as a massive intelligence failure. If we take network analysis seriously, then the failure leading to 9/11 was the result of not having built up human intelligence resources capable

---

[xiii] Duncan J. Watts, 1999. "Networks, Dynamics, and the Small-World Phenomenon," American Journal of Sociology, volume 13, number 2, pp. 493-527.

of recognizing and responding to the evolution of Al-Qaeda and its field operatives.[xiv] Michael Porter has been arguing for over two decades in competitive strategy that when firms compete with their buyers and suppliers, the more concentrated group wins, claiming the bulk of the profits for themselves.[xv] If I were the head of any nation's intelligence service, I would be asking myself very seriously why terrorist organizations are able to achieve a higher level of coordination and robustness than my own counter-terrorist division?

As compared to static models, Krebs' work analyzes the dynamics of the network and also recognizes the centrality measure's sensitivity to changes in nodes and links. In terms of utility as a counter-intelligence tool, the mapping exposes a concentration of links around the pilots, an organizational weakness which could have been used against the hijackers had the mapping been available prior to, rather than after the disaster.[xvi]

Because nodes with high centrality are potential points of failure they need to be mapped and monitored and whenever possible, removed. If enough nodes with high centrality are removed at the same time, this will cause the network to fragment into unconnected sub-networks. ***Naturally, the more compartmented the organization is, the fewer nodes of centrality need be removed*** in order to cause the network to implode. One trigger for moving from monitoring to disabling is a sudden increase in the flow of either money or information flow between links, and rapidly forming connections. In this regard, SIGINT can prove a powerful supplement to HUMINT. If an initial map of members and connections can be assembled, then SIGINT can indicate critical time periods. If we take Krebs' comment about the 9/11 network possessing self-organizing system properties, then ***the optimal time to intervene and remove high-centrality nodes is just at the beginning of the organizational equivalent of a "phase transition"***.[xvii]

### **Social Cohesion and Adhesion: Further Measures of Organizational Structure**

Moody and White provide an expansion of the social solidarity concept and the understanding of linkages between members of a community, the changing interconnections and the impact on node connectivity in "Social Cohesion and Embeddedness".[xviii] They argue that the defining characteristic of a strongly cohesive group is that "it has a status beyond any individual group member". The authors define structural cohesion as "the minimum number of actors who, if removed from a group,

---

[xiv] See Reuel Marc Gerecht, "The Counterterrorist Myth", The Atlantic Monthly, July-August, 2001.
[xv] See Michael Porter, (a) "How Competitive Forces Shape Strategy", Harvard Business Review, July-August, 1979, (b) "What is Strategy", Harvard Business Review, November-December, 1996.
[xvi] See Peter Klerks, "The Network Paradigm Applied to Criminal Organisations: Theoretical Nitpicking or a relevant doctrine for investigators? Recent developments for the Netherlands," Connections, 24(3): 53-65 http://www.sfu.ca/~insna/Connections-Web/Volume24-3/klerks.pdf
[xvii] For more detail on this subject see Michael Lissack, "Chaos and Complexity -- What does that have to do with knowledge management?" Knowledge Management: Organization, Competence and Methodology, ed. J. F. Schreinemakers. Wurzburg, Germany, Ergon Verlog. 1: 62-81. http://www.lissack.com/writings/knowledge.htm
[xviii] James Moody, Douglas R. White, "Social Cohesion and Embeddedness: A Hierarchical Conception of Social Groups, Santa Fe Institute Working Papers, 00-08-049, http://www.santafe.edu/sfi/publications/Working-Papers/00-08-049.pdf

would disconnect the group", leading to hierarchically nested groups, where highly cohesive groups are embedded within less cohesive groups. Thus, cohesions is an emergent property of the relational pattern that holds a group together.

As the dynamical process of group development unfolds, typically a weak form of structural cohesion begins to emerge as collections of unrelated individuals begin connecting through a single path which reflects new relationships. As additional relations form among previously connected pairs of individuals, multiple paths through the group develop, increasing the community's ability to "hold together".

In situations where relations revolve around a leader, the group is often described as "notoriously fragile", illustrating the fact that *increasing relational volume thru a single individual does not necessarily promote cohesiveness*. Nevertheless, groups with an all-in-one relational organization such as *terrorist networks, may be stable and robust to disruptions if "extraordinary efforts" are put into maintaining their weak relational structure*. The spoke-and-hub configuration of these networks thrives on the lack of knowledge that each particular node has about the organization as a whole, a captured or destroyed link in the network does not put the organization at risk. The stability of such groups depends on the ability to keep the hub hidden, because *the hub then becomes the entire group's fundamental structural weakness*.

Weakly cohesive organizations also promote segmentation into structures that are only minimally connected to the rest of the group, leading to schisms and factions. These organizations are also easily disrupted by individuals leaving the group. *Usually, individuals whose removal would disconnect the group are those in control of the flow of resources in the network*.

On the contrary, *collectivities that do not depend on individual actors are less easily segmented*. These highly cohesive groups benefit form the existence of multiple paths and sets of alternative linkages, with no individual or minority within the group exercising control over resources. The "multiple connectivity" is thus the essential feature of the strong structurally cohesive organizations.

An interesting characteristic of such highly cohesive networks (**HCN's**) is that they are characterized by a reduction in the power provided by structural holes, such that the ability of any individual to have power within the setting is limited as connectivity increases.[xix] For a structurally cohesive group, the information transmission increases with each additional independent path in the network, which may infer that high connectivity leads to more reliability as information is combined from independent multiple sources. "local pockets of high connectivity" can act as "amplifying substations" of information and/or resources. Moody and White relate this operationalization to the actors' relative involvement depth in social relations, as defined by the concept of *embeddedness*:[xx]

> "If cohesive groups are nested within each other, then each successive group is more deeply embedded within the network. As such, one aspect

---

[xix] In terms of strategy, destabilizing this kind of network means pressuring the group to increase its recruitment and raise its connectivity as opposed to the strategy of forced over-compartmentation. Induced excess connectivity represents a different kind of complexity overload.

[xx] Ibid. No. 17

of embeddedness- the depth of involvement in a relational structure- is captured by the extent to which a group is nested within the relational structure."

In a companion paper to Moody and White's "Social Cohesion and Embeddedness", White and Harary, in distinguishing between the adhesion concept related to the attractive or charismatic qualities of leaders (or attractions to their followers) that create weaker or stronger many-to-one ties or commitments, and the cohesion defined by the many-to-many ties among individuals, as they form into clusters.[xxi]

The authors reiterate the intuitive aspects of the cohesion's definition: a group is cohesive to the extent that the social relations of its members are resistant to the group being pulled apart, and a group is cohesive to the extent that the multiple social relations of its members pull it together. In revisiting the idea that *minimal cohesion occurs in social networks with a strong group leader or popular figure*, White and Harary introduce the concept of *"adherents"* of a social group to specify *"the many-to-one commitments of individuals to the group itself or to its leadership"*.

> "*What holds the group together where this is the major factor in group solidarity is the strength of adhesion of members to the leader, not the cohesiveness of group members in terms of social ties amongst themselves*. The model of "adhesion" rather than cohesion might apply to the case of a purely vertical bureaucracy where there are no lateral ties. "

*As a general definition, a group is adhesive to the extent that the social relations of its members are pairwise-resistant to being pulled apart.[xxii] Another element of group robustness is the redundancy of connections*:[xxiii]

> The level of cohesion is higher when members of a group are connected as opposed to disconnected, and further, when the group and its actors are not only connected but also have redundancies in their interconnections. The higher the redundancies of independent connections between pairs of nodes, the higher the cohesion, and the more social circles in which any pair of persons is contained.

---

[xxi] The Cohesiveness of Blocks in Social Networks: Node Connectivity and Conditional Density. Submitted to Sociological Methodology 2001. http://www.santafe.edu/files/workshops/dynamics/sm-wh8a.pdf

[xxii] The concept of cohesion is formalized through the use of graph theory. The graph is defined so that the vertices represent the set of individuals in the network, and the edges are the relations among actors defined as paired sets. The subsets of nodes that link non-adjacent vertices will disconnect actors if removed. Any such set of nodes is called an (i, j) cut-set if every path connecting i and j passes through at least one node of the set . The "cut-set resistance to being pulled apart" criterion and the multiple independent paths "held together" criterion of cohesion are formally equivalent in this formal specification. This kind of graph, if constructed with complete information, also provides a predictive mechanism for exactly which nodes need to be removed in order to remove the possibility of signals propagating through the system.

[xxiii] Ibid

The important consideration for counter-intelligence here is that ***the higher the level of redundancy, the more likely the existence of the group is to be revealed and the easier it is to create a map of social network relationships.*** A major part of the successful exploitation of this group characteristic is another basic counter-intelligence principle—coverage. Good coverage will yield good observations from which good social network maps can be derived. The caution here, as we have already noted, is like many other HUMINT activities, good coverage is not possible to achieve solely by satellite reconnaissance or any other national technical means (NTM). Of course, once target individuals have been identified, dedicated remote sensing technology can, in fact, be a very helpful adjunct to the processes of coverage, compartmentation and penetration.

Another logical inference is that ***measurable differences in cohesiveness should have predictive consequences*** for social groups and their members across many different social contexts. In terms of counter-terrorism, this is a clear, predictive model which can be utilized for the optimization of effort and resources. It's not merely a case of getting the most "bang for the buck", but it is an ideal mid-range solution for empirically validated scientific research to provide a typology which allows groups to be characterized in such a way that when resources are directed at preventing the group from executing terrorist attacks, the resources are used in an operationally efficient fashion. To bring it down to the micro level, ***it won't do any good to remove the leader of a group which is characterized by a strong measure of adhesion***. Just as ***it would do no good to attempt to counter a group with strong cohesion by removing its charismatic leader***. If no other lesson for counter terrorism comes out of this discussion, ***understanding this one critical rule is enough to save millions of dollars and thousands of lives***.